\documentclass[aps,prc,twocolumn,floatfix,showpacs,superscriptaddress,showpacs,amsfonts,amssymb]{revtex4}
\usepackage{amsmath}
\usepackage{graphicx}
\begin{document}
\title{Level density of 2$^{+}$ states in $^{40}$Ca from high energy-resolution ($p,p^{\prime}$) experiments}

\author{I.~Usman}
\affiliation{iThemba LABS, PO Box 722, Somerset West
7129, South Africa}\affiliation{School of Physics, University of the
Witwatersrand, Johannesburg 2050, South
Africa}
\author{Z.~Buthelezi}
\affiliation{iThemba LABS, PO Box 722, Somerset West 7129,
South Africa}
\author{J.~Carter}
\affiliation{School of Physics, University of the
Witwatersrand, Johannesburg 2050, South Africa}
\author{G.~R.~J.~Cooper}
\affiliation{School of Earth Sciences, University of the
Witwatersrand, Johannesburg 2050, South Africa}
\author{R.~W.~Fearick}
\affiliation{Department of Physics, University of Cape Town,
Rondebosch 7700, South Africa}
\author{S.~V.~F\"ortsch}
\affiliation{iThemba LABS, PO Box 722, Somerset West 7129,
South Africa}
\author{H.~Fujita}\altaffiliation[Present address:]{Research Center for Nuclear Physics, Osaka University, Ibaraki, Osaka 567-0047, Japan}
\affiliation{iThemba LABS, PO Box 722, Somerset West 7129,
South Africa}\affiliation{School of Physics, University of the
Witwatersrand, Johannesburg 2050, South Africa}
\author{Y.~Kalmykov}
\affiliation{Institut f\"ur Kernphysik, Technische
Universit\"at Darmstadt, D-64289, Darmstadt, Germany}
\author{P.~von~Neumann-Cosel}\email{vnc@ikp.tu-darmstadt.de}
\affiliation{Institut f\"ur Kernphysik, Technische
Universit\"at Darmstadt, D-64289, Darmstadt, Germany}
\author{R.~Neveling}
\affiliation{iThemba LABS, PO Box 722, Somerset West 7129,
South Africa}
\author{I.~Poltoratska}
\affiliation{Institut f\"ur Kernphysik, Technische
Universit\"at Darmstadt, D-64289, Darmstadt, Germany}
\author{A.~Richter}
\affiliation{Institut f\"ur Kernphysik, Technische
Universit\"at Darmstadt, D-64289, Darmstadt, Germany}
\affiliation{ECT$^{\ast}$, Villa Tambosi, I-38123 Villazzano
(Trento), Italy}
\author{A.~Shevchenko}
\affiliation{Institut f\"ur Kernphysik, Technische
Universit\"at Darmstadt, D-64289, Darmstadt, Germany}
\author{E.~Sideras-Haddad}
\affiliation{School of Physics, University of the
Witwatersrand, Johannesburg 2050, South Africa}
\author{F.~D.~Smit}
\affiliation{iThemba LABS, PO Box 722, Somerset West 7129,
South Africa}
\author{J.~Wambach}
\affiliation{Institut f\"ur Kernphysik, Technische
Universit\"at Darmstadt, D-64289, Darmstadt, Germany}

\date{\today}

\begin{abstract}
The level density of 2$^{+}$ states in $^{40}$Ca has been
extracted in the energy region of the isoscalar giant quadrupole resonance (ISGQR) from a fluctuation analysis of high energy-resolution ($p,p^{\prime}$) data taken at incident energies of 200 MeV at the K600 magnetic spectrometer of iThemba LABS, South Africa. Quasi-free scattering cross sections were calculated to estimate their role as a background contribution to the spectra and found to be small.
The shape of the background was determined from the discrete wavelet transform of the spectra using a biorthogonal wavelet function normalized at the lowest particle separation threshold. The experimental results are compared to widely used phenomenological and microscopic models.

\end{abstract}

\pacs{21.10.Ma, 25.40.Ep, 24.30.Cz, 27.40.+z}

\maketitle
\section{Introduction}

Level densities are fundamental quantities in the description of many-body systems \cite{hui72}. Besides their importance as a basic nuclear structure property it is well known that, through the statistical
model of nuclear reactions, level densities have a
strong impact on the results of calculations of other nuclear physics
observables. This is particularly so for thermonuclear rates in
nucleosynthesis models \cite{rau00,arn07}, in fission and fusion reactor
design \cite{cap09} as well as for the derivation of $\gamma$-strength functions from the decay of highly excited nuclei \cite{lar11}.

Experimental information on level densities is largely confined to low excitation energies, where knowledge of the excited states is rather complete, and just above the particle emission thresholds, where resonance spacings can be determined from capture reactions. Some information on level densities at higher excitation energies has been extracted from the analysis of Ericson fluctuations (see, e.g., Ref.~\cite{gri02} and references therein). Here we present results derived from high energy-resolution measurements of scattering cross sections in the energy region of giant resonances. The method is based on a fluctuation analysis of the spectra \cite{han90,mue83} and has been successfully applied recently to giant resonance data for a variety of modes like Gamow-Teller (GT) or electric and magnetic quadrupole resonances \cite{kal06,kal07}.
It does not provide a total but a spin- and parity-resolved level density determined by the quantum numbers of the investigated resonance.
In the present work we study 2$^{+}$ states in $^{40}$Ca from an analysis of high energy-resolution ($p,p^{\prime}$) data taken at an incident energy of 200 MeV and for kinematics favoring population of the ISGQR at the K600 magnetic spectrometer of iThemba LABS, South Africa \cite{nev11}.

A prerequisite of the method is the decomposition of the spectra into the part stemming from excitation of the ISGQR and any background. The latter may contain contributions from the
experimental instrumentation, excitation of other multipoles, and other physical processes. For hadronic probes quasi-free scattering must be considered as a source of background in the energy region above the lowest particle threshold, where the giant resonance strength resides.
For the case of proton scattering considered here, models have been developed in the framework of the Distorted Wave Impulse Approximation (DWIA) and have been shown to be quite successful at incident energies up to a few hundred MeV \cite{bak97}. Alternatively, the shape of the underlying background can be determined in a largely model-independent way using a discrete wavelet analysis \cite{kal06}.

The results will be compared to theoretical level densities from approaches based on the phenomenological back-shifted Fermi gas model (BSFG) \cite{rau97,egi05} and microscopic Harteee-Fock-Bardeen-Cooper-Schrieffer (HF-BCS) \cite{dem01} and Hartree-Fock-Bogolyubov (HFB) \cite{gor08} calculations. Experimental tests of the latter are of particular interest since the predictive power of the BSFG for extrapolations to  exotic nuclei is limited. Thus, network calculations of the astrophysical $r$-process typically  depend on microscopic level densities \cite{arn07}. Nuclei with shell closures are particularly difficult to describe in the phenomenological models because their parameters do not follow the smooth systematics as a function of basic quantities like mass number observed otherwise, and $^{40}$Ca is thus special because of its doubly magic nature. In addition, because of the weakening of low-energy quadrupole vibrations in doubly magic nuclei it also provides a test of vibrational enhancement factors in the microscopic calculations.

\section{Experiment}\label{sec:experiment}

The fine structure of the ISGQR in medium-mass and heavy nuclei has been investigated with high-resolution $(p,p')$ scattering at iThemba LABS with the aim to extract
information about their dominant decay processes \cite{she04,she09}. Recently, these studies have been extended
to the low-mass region $12 \le A \le 40$ \cite{usm09,usm11}. The
present work focuses on the case of $^{40}$Ca where highly fragmented $E2$ strength has been observed with a variety of probes including inelastic electron \cite{die94}, proton \cite{sch01} and $\alpha$
\cite{bor81} scattering.

\begin{figure}[tbh] 
\includegraphics[width=8.6cm]{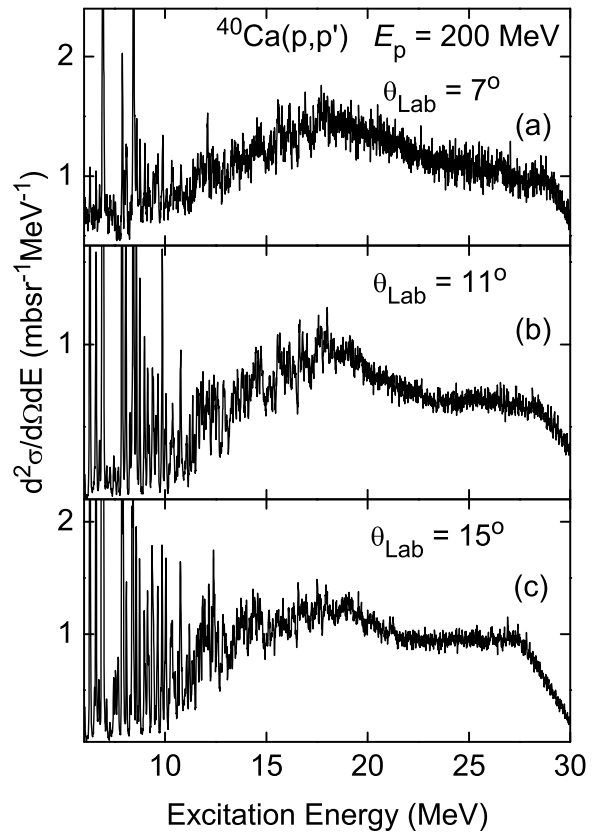} 
 \caption{Excitation energy spectra for $^{40}$Ca for $E_x \approx 6 - 30$ MeV at
scattering angles $\theta_{\rm{Lab}} = 7^{\circ}$,
11$^{\circ}$ and 15$^{\circ}$. Note that the ISGQR is expected to be most strongly excited at $\theta_{\rm{Lab}}$ = 11$^{\circ}$. See also \cite{usm11}.}
\label{fig:targetspectra}
\end{figure}

In the experiment a 200 MeV proton beam
produced by the Separated Sector Cyclotron of iThemba LABS was inelastically scattered off a natural Ca target (areal density  3.0 mg/cm$^{2}$) and detected with the K600 magnetic spectrometer. In order to achieve high energy resolution the beam dispersion was matched to the spectrometer leading to a resolution $\Delta$\emph{E} = 35 - 40 keV (full width at half
maximum, FWHM). Data were taken at scattering angles $\theta_{Lab}$ = 7$^{\circ}$, 11$^{\circ}$ and 15$^{\circ}$ chosen
to lie below, at, and above the maximum of the cross
sections for $\Delta L = 2$ transitions populating the ISGQR.
The momentum acceptance of the spectrometer allowed to cover excitation energies between 6 and 30 MeV in $^{40}$Ca with a single field setting. Details of the data analysis are described elsewhere \cite{usm09}. The resulting spectra are displayed in Fig.~\ref{fig:targetspectra}.

The spectra, already discussed in Ref.~\cite{usm11}, exhibit a broad bump (roughly between 12 and 22 MeV) peaking around 18 MeV for scattering angles $\theta_{Lab} =  11^{\circ}$ and 15$^{\circ}$. It is associated with excitation of the ISGQR. Below $E_x \simeq 10$ MeV, many strong discrete transitions are visible with varying angular momentum transfer. Pronounced fine structure is visible in the excitation region of the ISGQR up to about 20 MeV.
The intermediate structure with peaks around 12, 14,
16, 17 and 18 MeV is consistent with previous experimental results
\cite{die94,sch01,bor81}.  At the smaller
scattering angle $\theta_{\rm{Lab}}$ = 7$^{\circ}$, the maximum cross section is shifted to about 17 MeV, and the fine
structure changes considerably indicating the presence of other
multipoles. Arguments for the predominance of quadrupole excitations in the $11^{\circ}$ and 15$^{\circ}$ spectra including a cross-correlation analysis have been presented in Ref.~\cite{usm11}.

\section{Background determination}\label{sec:background}

As pointed out above, knowledge of background in the spectra not related to population of the ISGQR is mandatory for an extraction of the level densities. The background can have contributions from
other multipoles excited, quasi-free scattering and from the
experimental instrumentation. The latter has been shown to be negligible in the present case \cite{usm09}. Quasi-free scattering dominates the cross sections in intermediate-energy proton scattering at excitation energies above the giant-resonance region \cite{bak97} and could thus to contribute substantially to the data investigated here. Their relevance is estimated from DWIA model calculations, which have been validated against a variety of data for incident proton energies up to 200 MeV (see, e.g., Refs.~\cite{cow89,car01,nev02}). Alternatively, the shape of the underlying background can be determined in a largely model-independent way using a wavelet analysis of the discrete wavelet transform (DWT) of the spectra \cite{kal06}. This method has been successfully applied \cite{kal07} to proton scattering spectra of $^{58}$Ni and $^{90}$Zr in the energy region of the ISGQR \cite{she09}.

\subsection{Quasi-free scattering calculations}\label{sec:quasifree}
The code THREEDEE \cite{cha98} was used for
the calculation of quasi-free nucleon knockout contributions from
the reactions $^{40}$Ca($p,2p$)$^{39}$K and
$^{40}$Ca($p,pn$)$^{39}$Ca at an incident proton energy of 200
MeV. The DWIA is
used to determine the contribution due to quasi-free proton and
neutron knockout in the inclusive proton-scattering reaction based on the assumption of a simple quasi-free
projectile-nucleon interaction. Such reactions indeed contribute to
the background underlying the isoscalar giant quadrupole
resonance in $^{40}$Ca as demonstrated in
earlier studies of
$^{40}$Ca($p,p^\prime\rm{p}$), $^{48}$Ca($p,p^\prime n$)
and $^{40}$Ca($p,p^\prime\alpha$)
angular correlations \cite{sch01,car01}.

The low-energy proton and neutron optical-potential parameters used
for the description of the distorted outgoing waves can be
found in Ref.~\cite{bec69}. Optical potential parameters used
for generating distorted waves of the high-energy incoming
proton and its subsequent quasi-free scattering outgoing proton
stem from the energy- and mass-dependent parametrization of Ref.~\cite{sch82} extrapolated to energies up to 200 MeV.  The Woods-Saxon
well radius and diffuseness parameters for the calculation of
proton and neutron bound-state wave functions were taken from Elton
and Swift \cite{elt67}. In order to perform the integration
over the kinematics of the knocked-out particle, the recoil momentum was chosen to be less than 200 MeV/c to determine the range
of the quasi-free scattering primary angles at each
proton energy. The program QUASTA \cite{quas} was used for the kinematic calculations. Contributions to the cross sections
due to the quasi-free process were obtained in terms of the sum
of the cross sections for knockout from the $1d_{3/2}$,
$2s_{1/2}$ and $1d_{5/2}$ states for both protons and neutrons
with spectroscopic factors taken from Ref.~\cite{ant81}
for proton and Refs.~\cite{wat82,ahm84} for neutron
states, except for the neutron
$1d_{3/2}$ shell where the shell-model limit of 4.0 was used.
Calculations were performed within a range of excitation
energies $E_x = 10 - 30$ MeV  corresponding to ejectile energies $E_{p}^{\prime} \simeq 190 - 170$ MeV, respectively.
In previous work at lower incident energy it was found that the
contribution due to the $^{40}$Ca($p,p^{\prime}\alpha$) reaction was very small \cite{sch01,car01}. Test calculations confirmed this for the present case and, therefore, quasi-free
($p,p^{\prime}\alpha$) scattering was neglected.
\begin{figure}[htb!]
 \includegraphics[width=8.6cm]{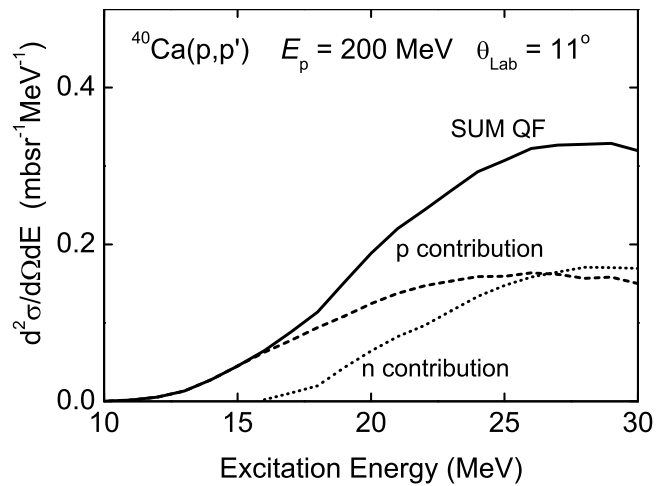}
\caption{Results of the DWIA
calculations for quasi-free reaction cross sections (SUM QF) of the  $^{40}$Ca($p,p^{\prime}$)reaction at
$E_p = 200$ MeV and $\theta_{Lab} = 11^{\circ}$ (solid line)  and their decomposition into contributions from $(p,2p)$ (dashed line) and $(p,pn)$ (dotted line) reactions.}
\label{fig:pncontributions}
\end{figure}

\begin{figure}[htb!]
  \includegraphics[width=8.6cm]{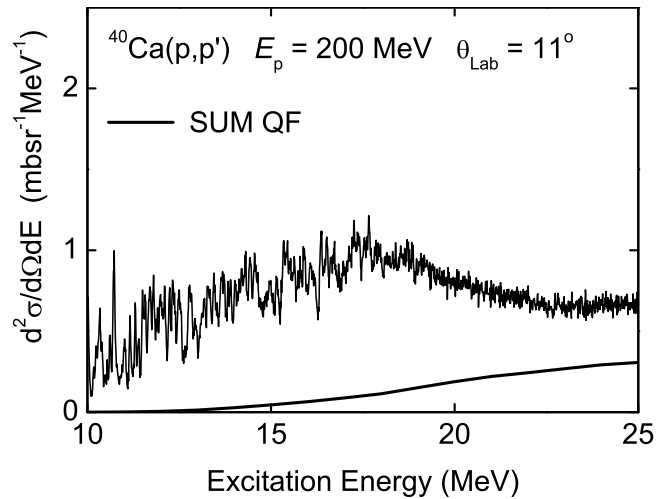}
  \caption{Quasi-free background contribution (SUM QF) to the spectrum of the  $^{40}$Ca($p,p^{\prime}$)reaction at
$E_p = 200$ MeV and $\theta_{Lab} = 11^{\circ}$.}
  \label{fig:qfbkgrd11}
\end{figure}

Results of the calculations are shown in
Fig.~\ref{fig:pncontributions} for $\theta_{Lab} = 11^\circ$ by way of example. Because of the much lower proton emission threshold ($S_p = 8.3$ MeV), the $(p,2p)$ reaction dominates up to $E_x \approx 25$ MeV, while contributions from the $(p,pn)$ reaction are comparable at larger excitation energies but small in the energy region of the ISGQR because of the high neutron separation energy ($S_n = 15.6$ MeV).
In Fig.~\ref{fig:qfbkgrd11}, the predicted total quasi-free cross section is compared with the measured $^{40}$Ca($p,p^{\prime}$) cross sections. It can be seen that the quasi-free
contribution is relevant only for excitation energies $E_x > 20$  MeV and even at 25 MeV it represents not more than about half of the experimental value. The quasi-free parts of the cross sections at other scattering angles are comparable \cite{usm09}. Clearly, the quasi-free process contributes little in the excitation energy region of interest and thus cannot provide a good approximation of the background.

\subsection{Discrete Wavelet Transform}\label{sec:dwt}

In high-energy resolution spectra dominated by a single giant resonance, the shape of the underlying background can be determined in a model-independent way using a wavelet analysis of the DWT of the spectrum \cite{kal06}. A comprehensive description of the application of wavelet analysis to high-resolution nuclear spectra can be found in Ref.~\cite{she08}. Here, we only reiterate a few basic facts necessary to understand the specific approach to background determination.
The wavelet analysis is performed by folding the original spectrum $\sigma (E)$ with a wavelet function $\Psi$, resulting in wavelet coefficients
\begin{equation}
    \label {eq:coeff} C(E_x, \delta E) = \frac{1}{\sqrt{\delta E}}
        \int \sigma(E) \Psi \left( \frac{E_x - E}{\delta E} \right) dE.
\end{equation}
The parameters (excitation energy $E_{x}$ and scale
$\delta E$) can be varied in continuous (adjustable to the specfic problem) or discrete ($\delta E= 2^{j}$, $E_{x}=
k\delta E$, $j,k = 1,2,3 \dots$) steps leading to a continuous wavelet transform (CWT) or DWT of the original spectrum, respectively.  The CWT can be used to extract scales characterizing the fine structure of the spectra \cite{she04,she09,usm11,pet10}. The DWT, while limited in scale resolution, allows to reassemble the
original signal from the wavelet coefficients.

\begin{figure}[htb!]
 \includegraphics[width=7cm]{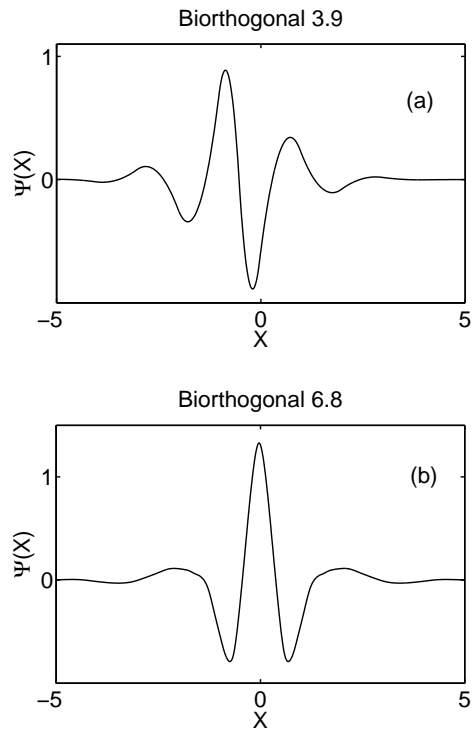}
  \caption{Examples of wavelet functions used in the present analysis. (a) BIOR3.9 wavelet. (b) BIOR6.8 wavelet.}
  \label{fig:bior-wavelets}
\end{figure}

The present
application furthermore utilizes the property of vanishing moments
fulfilled by many wavelet functions, namely
\begin{equation}
   \label{eq:vanishingm}
   \int {E^n \Psi \left( E \right)dE =
   0,\;\; n = 0,1...m}.
\end{equation}
When Eq.~(\ref{eq:vanishingm}) holds, any nonresonant background in the spectrum, whether of physical or experimental nature, does not
contribute to the wavelet coefficients as long as it can be approximated by a polynomial function of order $m$.
In the present analysis the biorthogonal family of mother
wavelets BIOR$Nr.Nd$ was used, where $Nr$ indicates the $n^{th}$ vanishing moment with a polynomial
function of order $n-1$ while $Nd$ represents the level of
decomposition \cite{mat06}. Examples of BIOR wavelets are shown in Fig.~\ref{fig:bior-wavelets}.
\begin{figure*}[htb!]
 \centering{
 \includegraphics[height=15cm,angle=-90]{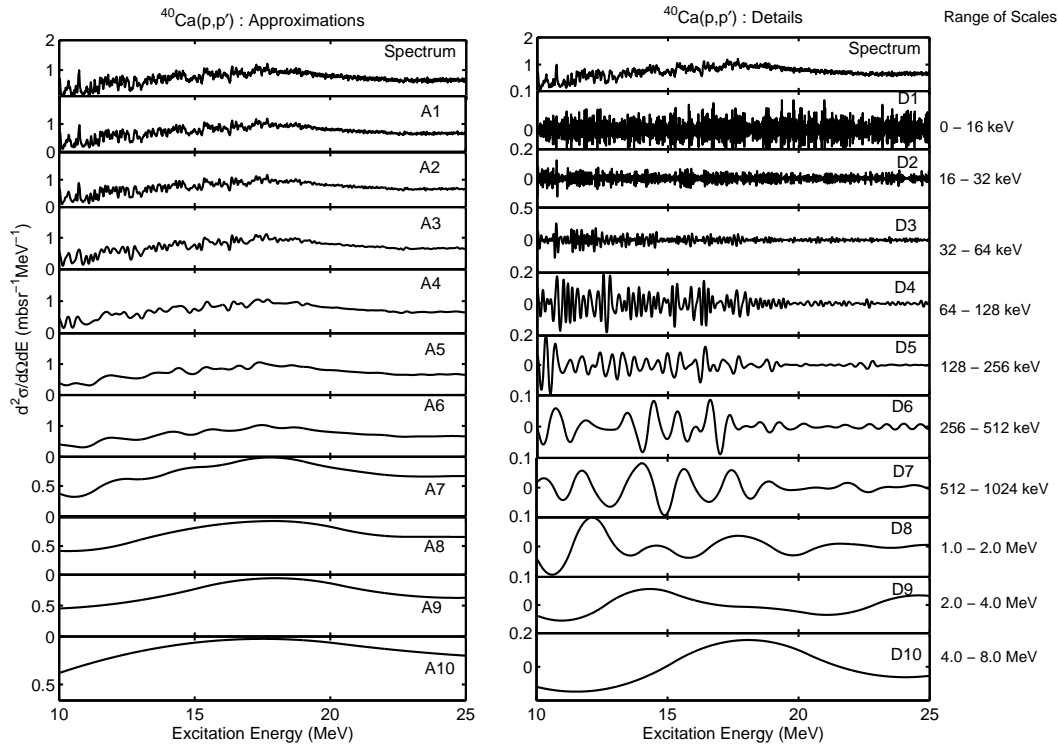}
\caption{DWT decomposition of the excitation energy spectrum of the $^{40}$Ca($p,p^{\prime}$) reaction measured at $E_p = 200$ MeV for $\theta_{Lab} = 11^\circ$ into approximations $Ai$ and details $Di$.}
\label{fig:dwtcs}
}
\end{figure*}

The DWT can be viewed as an iterative
decomposition in the form of low-pass and high-pass filtering
of the data into a sequence of approximations ($A_{j}$) and
details ($D_{j}$) with increasing scale $\delta E_j$. At each level of decomposition $A_{j} + D_{j} = A_{j-1}$. The approximations are the filtered
signals which provide the non-resonant background at the given range of scale values while the
details obtained from the wavelet coefficients provide the part of
the signal that was removed in the filtering process at that
level. Application of the DWT to the spectrum measured at $\theta_{Lab} = 11^\circ$ is shown in Fig.~\ref{fig:dwtcs}. The distributions of details indicate that scales between about 100 keV and 1 MeV ($D_4 - D_7$) contribute most to the fine structure in the energy region of the ISGQR. This is fully consistent with a wavelet analysis based on CWT  \cite{usm11} which finds maxima of the power spectrum of the wavelet coefficients (so-called characteristic scales) at values 150 keV, 240 keV, 460 keV, and 1.05 MeV.

\begin{figure}[htb!]
 \includegraphics[width=8.6cm,angle=0]{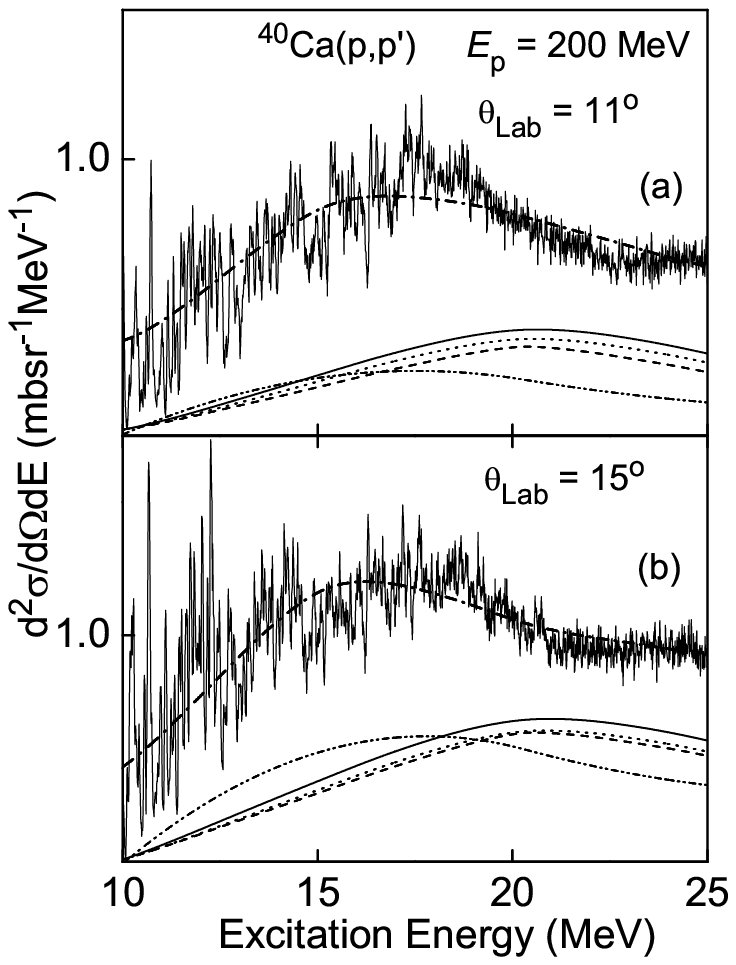}
\caption{Comparison of the spectrum of the $^{40}$Ca($p,p^{\prime}$) reaction measured at $E_p = 200$ MeV for (a) $\theta_{Lab} = 11^\circ$  and (b) $15^\circ$ to approximation $A9$ (dashed-dotted lines) determined from the DWT (cf.\ Fig.~\ref{fig:dwtcs}). The shape of the ISGQR is well reproduced.
The resulting background (approximation $A10$) is shown for different biorthogonal wavelet functions: BIOR6.8 (solid lines), BIOR5.5 (dotted lines), BIOR4.4 (dashed lines), and BIOR3.9 (dashed-doubly dotted lines).
}
\label{fig:biorbkgrds}
\end{figure}

At some level of decomposition the largest physical scale in the spectrum, viz.\ the width of the ISGQR is reached in the approximations ($A9$ in the present example). This is demonstrated in Fig.~\ref{fig:biorbkgrds}, where the the spectra measured at $11^\circ$ and $15^\circ$ are compared to the approximations A9 of the respective DWT analysis shown as dashed-dotted lines. The main structure between 10 and 20 MeV associated with the ISGQR is well described in both cases. Structures at even larger scales (see approximation $A10$ in Fig.~\ref{fig:dwtcs}) are thus associated with the background.

The impact of a variation of the number of vanishing moments in the DWT analysis is also demonstrated in Fig.~\ref{fig:biorbkgrds}, where the background shapes $A$10 deduced with different biorthogonal wavelet functions are compared. While the energy dependence extracted with BIOR3.9 differs, the spectral shapes are very similar for larger numbers of vanishing moments. The results discussed in the following are based on the application of a BIOR6.8 wavelet. A fine tuning of the resulting background is then carried out
by shifting of the $A$10 form in vertical direction in order to
satisfy the experimental condition that the spectrum is background-free below the proton threshold.

\section{Level density of 2$^{+}$ states}\label{leveldensity}

\subsection{Fluctuation analysis}
\begin{figure}[tbh]
 \centering
 \includegraphics[width=8.6cm]{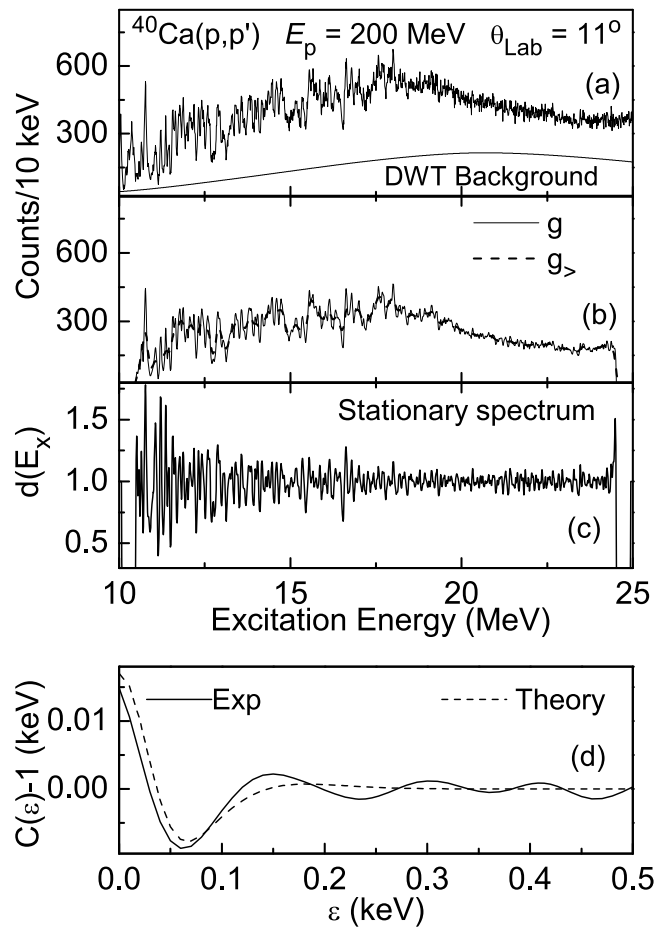}
\centering
\caption{Illustration of the autocorrelation analysis: (a) Experimental $^{40}$Ca($p,p^{\prime}$) spectrum including background obtained with the use of discrete wavelet transform, (b) background-subtracted smoothed spectra \emph{g}($E_{\rm{x}}$) and $g_{>}$($E_{\rm{x}}$), respectively, (c) stationary fluctuating spectrum d$\left( E_{\rm{x}}\right) $ obtained by dividing the two smoothed spectra, and (d) autocorrelation functions from experiment and Eq.~(\ref{eq:analytical}) shown as solid and dashed line, respectively.}
\label{fig:autocorrelation}
\end{figure}

In this section, the level density of 2$^{+}$ states in
$^{40}$Ca extracted by means of a self-consistent procedure
based on a fluctuation analysis \cite{han90} in the excitation energy
interval between 10 and 20 MeV is discussed. As a starting
point the measured excitation energy spectrum for
$^{40}$Ca($p,p^{\prime}$) at $\theta_{Lab} = 11^\circ$
together with the background deduced from the DWT analysis with a BIOR6.8 wavelet (solid line) is shown in Fig.~\ref{fig:autocorrelation}(a).
The method
discussed here is applicable in a region where
${\langle{\Gamma}\rangle}/{\langle{D}\rangle} \ll$ 1 but
$\langle{D}\rangle < \Delta{E}$. Here,
$\langle{\Gamma}\rangle$ represents the mean level width,
$\langle{D}\rangle$ is the mean level spacing and $\Delta{E}$
is the experimental energy resolution. The first step of the
fluctuation analysis is a subtraction of the nonresonant background determined as described in the previous section  from the spectrum. Then a smoothing is performed using a Gaussian function with a width $\sigma$ smaller than the experimental energy resolution
$\Delta E$ in order to suppress contributions due to statistical fluctuations. The resulting spectrum is called $g\left(
E_{\rm{x}}\right)$. This spectrum is again folded with a
Gaussian, of a width $\sigma_{>}$ larger than the
experimental energy resolution, in order to remove gross
structures. The resulting spectrum referred
to as $g_{>}\left( E_{\rm{x}}\right)$ is displayed in Fig.~\ref{fig:autocorrelation}(b) and defines a mean value
around which the original data fluctuate. The ratio of $g\left(
E_{\rm{x}}\right)$ and $g_{>}\left( E_{\rm{x}}\right)$ is
called the stationary
spectrum which fluctuates around unity as demonstrated in Fig.~\ref{fig:autocorrelation}(c). It represents a direct measure of the local intensity
fluctuations which can be expressed in terms of an
autocorrelation function of the spectrum
\begin{equation}
 C\left(\epsilon\right)=\langle{d\left( E_{\rm{x}}\right)d\left( E_{\rm{x}}+\epsilon\right)}\rangle,
\end{equation}
where $\epsilon$ denotes the energy shift and the brackets $\langle \rangle$ indicate averaging over a suitable energy interval.
The value ($C\left(\epsilon = 0\right)- 1$) is the variance.

This experimental
autocorrelation function can be approximated by an analytical
expression \cite{han90}
\begin{equation}\label{eq:analytical}
 C\left(\epsilon\right) - 1 = \dfrac{\alpha\langle{D}\rangle}{2\Delta{E}\sqrt{\pi}}  f\left( \epsilon,\Delta{E},\sigma,\sigma_> \right),
\end{equation}
where the function $f$ is
normalized such that $f( \epsilon = 0) = 1$. From
Eq.~(\ref{eq:analytical}) it follows that the value of the autocorrelation
function ($C\left( \epsilon \right) - 1$) at $\epsilon = 0$,
i.e., the variance of the stationary spectrum, is
proportional to the mean level spacing $\langle{D}\rangle$.
Thus, $\langle{D}\rangle$ can be extracted directly when the normalized variance $\alpha$ of the underlying spectral distributions is known. It is assumed that because of the high excitation energies these can be approximated by the predictions of random matrix theory \cite{wei09}, i.e., a Wigner distribution for the level spacing and a Porter-Thomas (PT) distribution for the intensities. If one assumes that the cross sections result from a single class of states, i.e.\ $J^\pi = 2^+$ states in the region of the ISGQR, then $\alpha = \alpha_{Wigner} + \alpha_{PT} = 2.237$.
Figure~\ref{fig:autocorrelation}(d) illustrates the
autocorrelation functions for both the experimental data and the model. Since the $\epsilon$ dependence is contained in the function $f$, which depends on experimental parameters only, $\langle{D}\rangle$ is determined by the value of the autocorrelation function at $\epsilon = 0$.

\subsection{Experimental results}

The analysis requires a suitable interval length in order to keep finite-range-of-data errors \cite{ric74} at an acceptable level. In the present case 2 MeV intervals were chosen. The resulting level densities for the spectra measured at $11^\circ$ and $15^\circ$ are summarized in Tab.~\ref{table:leveldensity}. The results should be independent of the kinematics of the measurement and indeed agreement between the values deduced from the two spectra within experimental uncertainties is obtained, confirming the correctness of the assumptions underlying their extraction.
\begin{table}
\caption{\label{table:leveldensity}
Level density of 2$^{+}$ states in $^{40}$Ca at various excitation energies $E_x$ extracted from the ($p,p^{\prime}$) data at scattering angles $\theta_{Lab} = 11^{\circ}$ and 15 $^{\circ}$.}
\begin{ruledtabular}
\begin{tabular}{ccc}
$E_x$ (MeV) & \multicolumn{2}{c}{Level density (MeV$^{-1}$)}\\
& 11$^\circ$ & 15$^\circ$ \\
\hline
11 & $17.5^{+2.9}_{-3.1}$ & $18.1^{+3.0}_{-1.9}$ \\
13 & $64.1^{+10.6}_{-10.7}$ & $57.3^{+9.5}_{-6.6}$  \\
15 & $213^{+35}_{-38}$ & $ 224^{+37}_{-32}$\\
17 & $217^{+36}_{-43}$ & $ 236^{+39}_{-37}$ \\
19 & $1007^{+166}_{-226}$ & $705^{+116}_{-122}$ \\
\end{tabular}
\end{ruledtabular}
\end{table}
The uncertainties of the extracted level
densities given in Tab.~\ref{table:leveldensity} were obtained by varying the input parameters of
the autocorrelation function and the fluctuation analysis as well
as the background subtraction method and repeating the analysis.
The following contributions  assumed to be independent of each other have been considered in the calculation of the experimental error bars \\
(i) smoothing parameters: a variation by $\pm 10$\% leads to a 2\% error contribution. \\
(ii) Range of the excitation energy interval: the analysis  was repeated for interval sizes between 0.5 and 2 MeV. The corresponding uncertainty amounts to $\pm 10$\%. \\
(iii) Variations of the normalized variances of the spacing and intensity distributions in Eq.~(\ref{eq:analytical}): this is due to admixtures of states with another spin and/or parity and depends on the ratio of level densities and cross sections \cite{kil87}.  Asymptotically, for the spacing distribution one approaches a value $\alpha_s = 2 + 3 N_1/N_2$, where $N_1/N_2$ denotes the cross section ratio of the two multipoles.  An upper limit $N_1/N_2 = 0.1$ was assumed based on the arguments for a dominant excitation of the ISGQR in the spectra discussed in Ref.~\cite{usm11}. Variations in the variance of the intensity contribution can be neglected for small admixtures. Since $\alpha$ in Eq.~(\ref{eq:analytical}) only increases for values $N_1/N_2 \neq 0$, the level density also increases, leading to a $+13$\% error. \\
(iv) Choice of wavelet functions: the variation between the different choices of the BIOR wavelets (Fig.~\ref{fig:biorbkgrds}) is taken as an estimate of the uncertainty of the background determination. Increasing the number of vanishing moments leads to larger backgrounds but approaching constant magnitude and shape for the highest values used in the present analysis. The variation thus leads to a systematic reduction of the level densities reaching $-17$\% for the $11^\circ$ and $-13$\% for the $15^\circ$ spectrum, respectively.

\subsection{Model comparison}

\begin{figure}[tbh]
  \centering
 \includegraphics[width=8.6cm]{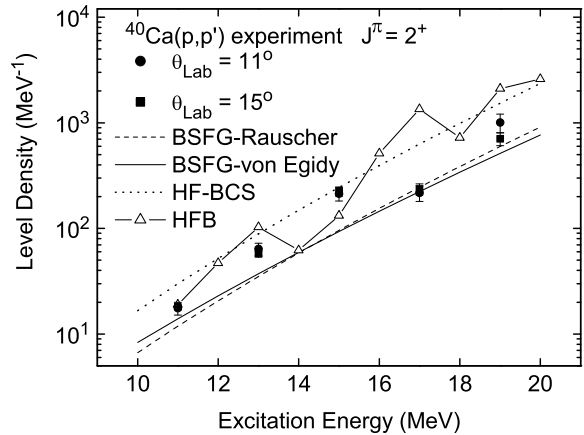}
  \centering
  \caption{Level density of 2$^{+}$ states in $^{40}$Ca extracted from the ($p,p^{\prime}$) data (full symbols) compared to model predictions: BSFG-Rauscher \cite{rau97} (dashed line),
  BSFG-von Egidy \cite{egi09} (solid line), HF-BCS \cite{dem01} (dotted line) and HFB \cite{gor08} (triangles, the connecting line is to guide the eye only).}
  \label{fig:dwtld}
\end{figure}

Shown in Fig.~\ref{fig:dwtld} are the experimental level densities deduced from both spectra in comparison to model calculations. The theoretical results considered include the phenomenological backshifted Fermi gas (BSFG) and microscopic models. Two different sets of values were taken for the BSFG parameters $\Delta$, the ground-state energy
correction accounting for pairing and shell effects, and the level density parameter $a$ describing the exponential increase with energy.
Rauscher {\it et al}.~\cite{rau97} provide a fit to stable nuclei across the nuclear chart which is used for astrophysical network calculations of the $s$-process, including extra parameters for an improvement of the description in local mass areas. Von Egidy and Bucurescu \cite{egi05} performed a global fit with parameters dependent on experimental masses only. The latter approach has been recently improved by a modification of the spin-cutoff parameter \cite{egi09}, which is in accordance with shell-model Monte Carlo calculations \cite{alh07}. A current microscopic approach is based on a Hartree-Fock Bogoliubov (HFB) plus combinatorial model \cite{hil06}. This has been been improved in Ref.~\cite{gor08} to include beyond rotational also vibrational degrees-of-freedom.
We also show a comparison with HF-BCS model results \cite{dem01}.

Both BSFG parametrizations provide very similar results for $2^+$ states in $^{40}$Ca in the  excitation energy range considered $E_x \simeq 10 - 20$ MeV. The corresponding level density parameter $a \approx 5.3$ MeV$^{-1}$, which is unusually low because of the double shell closure, provides a reasonable description of the energy dependence in the experimental results, but the magnitudes are about a factor of two too low. The energy dependence of the HF-BCS calculation is again very similar to the data and the BSFG results, but the predicted level densities are about 50\% too high. Since an implicit assumption of all three models is the equipartition of states with positive and negative parity for a given spin, the theoretical results were divided by a factor of two for the comparison in Fig.~\ref{fig:dwtld}.

The HFB model is capable of calculating level densities of states with given spin and parity and furthermore allows for deviations from the smooth energy dependence encoded within the BSFG and HF-BCS models by  the microscopically generated distribution of single-particle states and collective enhancement factors. The data indeed indicate such fluctuations, similar to findings in Ref.~\cite{kal07}. However, in detail the correspondence of fluctuations around the average increase of the level densities extracted from the data and the HFB calculations is limited. For example, at $E_x = 17$ MeV the data find a local minimum while the HFB result predicts a maximum. However, the absolute magnitude is reasonably described by the HFB calculations despite the fact that no renormalization (cf.\ Eq.~(9) of Ref.~\cite{gor08}) to the experimental level scheme at low energy and neutron resonance spacings was included. The average energy dependence describes the data up to 15 MeV well but overpredicts the experimental results at higher excitation energies.

\section{Conclusions and outlook}

The present work reports on the level density of $2^+$ states in $^{40}$Ca in the energy region of the ISGQR. It is extracted from high-energy resolution proton scattering spectra measured in kinematics favoring quadrupole transitions. The method is based on the analysis of cross-section fluctuations and thus is dependent on the highest possible energy resolution. The magnitude of the fluctuations is determined by an autocorrelation function and can be related to the average level spacing \cite{han90}.

A crucial input into the analysis is the knowledge of background contributions to the measured excitation energy spectra. In the present work, these were determined by a decomposition of the spectrum with a discrete wavelet analysis described in Ref.~\cite{she08} which was, for example, successfully applied \cite{kal07} to experiments studying the ISGQR \cite{she04,she09} and magnetic quadrupole resonance \cite{vnc99} in medium-heavy nuclei. Alternatively, the role of quasi-free reactions as a background source was estimated but found to be small. The DWT analysis is largely model-independent but depends on two assumptions, {\it viz.}\ that the energy dependence of background contributions in the spectra can be approximated by a polynomial and that one excitation mode dominates the spectra. Based on the agreement of experimental level densities deduced for different kinematics we conclude that these conditions are well fulfilled for the present case (for the latter point see also the discussion in Ref.~\cite{usm11}). The extracted level densities were compared to phenomenological BSFG and microscopic HF-BCS and HFB calculations. The excitation energy dependence is reasonably well described by the BSFG and HF-BCS models but absolute values are either over- or underestimated. The HFB calculation provides the correct magnitude up to about 15 MeV but overestimates the experimental values at higher excitation energies.

The relevance of this new technique combining a fluctuation analysis of high-energy resolution spectra with a discrete wavelet decomposition to determine background components is twofold: since the method works best in the energy region where giant resonances dominate the cross sections, it provides experimental data on level densities for $E_x \approx 10- 20$ MeV difficult to experimentally access otherwise, and it also provides information for states of specific spin and parity, thus allowing tests of the spin distribution models \cite{egi09} and claims of a parity dependence of level densities \cite{alh00} with significant astrophysical consequences \cite{loe08,hut11}. Beyond the application to electric and magnetic quadrupole modes discussed so far, new facilities allowing for high-resolution $0^\circ$ scattering experiments of hadronic beams \cite{nev11,tam09} promise data for $\Delta L = 0$ and 1 natural-parity modes from $\alpha$ scattering \cite{bra83,lu86} and proton scattering \cite{tam11}, respectively. Unnatural-parity states can be studied in backward-angle electron scattering \cite{vnc99,lue95} and charge-exchange reactions \cite{kal06,fuj07}. Systematic studies of spin- and parity-resolved level densities  along these lines are in progress.

\begin{acknowledgments}
We are indebted to J.~L.~Conradie and the accelerator crew at iThemba LABS for providing excellent proton beams. Useful discussions with S.~Goriely are acknowledged. This work has been supported by the National Research Foundation (South Africa) and by the Deutsche Forschungsgemeinschaft under contracts SFB 634 and NE 679/2-2.
\end{acknowledgments}

\end{document}